%% file: main.tex
\documentclass[runningheads]{llncs}
\usepackage{hyperref}
\usepackage{graphicx}
\usepackage{changes}
\usepackage{multirow}
\usepackage{enumitem}
\usepackage{algorithm}
\usepackage{algpseudocode}
\usepackage[export]{adjustbox}
\usepackage{amsmath}

\newcommand{\numberParticipants}{39 }
\newcommand{\reviewcolor}{black}

\begin{document}
\title{Exploring Natural Language Processing Methods for Interactive Behaviour Modelling}
\titlerunning{Exploring NLP Methods for Interactive Behaviour Modelling}
\author{
Guanhua Zhang$^1$\and
Matteo Bortoletto$^1$\and
Zhiming Hu$^{1,2}$\thanks{Corresponding author}\and
Lei Shi$^1$ \and
Mihai Bâce$^1$ \and
Andreas Bulling$^1$
}
\authorrunning{G. Zhang et al.}
\institute{$^1$Institute for Visualisation and Interactive Systems, $^2$Institute for Modelling and Simulation of Biomechanical Systems, University of Stuttgart, Stuttgart, Germany\\
\email{\{guanhua.zhang, matteo.bortoletto, zhiming.hu, lei.shi, mihai.bace, andreas.bulling\}@vis.uni-stuttgart.de}}
\maketitle              %
\begin{abstract} %
Analysing and modelling interactive behaviour is an important topic in human-computer interaction (HCI) and a key requirement for the development of intelligent interactive systems.
Interactive behaviour has a sequential (actions happen one after another) and hierarchical (a sequence of actions forms an activity driven by interaction goals) structure, which may be similar to the structure of natural language.
Designed based on such a structure, natural language processing (NLP) methods have achieved groundbreaking success in various downstream tasks.
However, few works linked interactive behaviour with natural language. %
In this paper, we explore the similarity between interactive behaviour and natural language by applying an NLP method, byte pair encoding (BPE), to encode mouse and keyboard behaviour.
We then analyse the vocabulary, i.e., the set of action sequences, learnt by BPE, as well as use the vocabulary to encode the input behaviour for interactive task recognition.
An existing dataset collected in constrained lab settings and our novel out-of-the-lab dataset were used for evaluation.
Results show that this natural language-inspired approach not only learns action sequences that reflect specific interaction goals, but also achieves higher F1 scores on task recognition than other methods.
Our work reveals the similarity between interactive behaviour and natural language, and presents the potential of applying the new pack of methods that leverage insights from NLP to model interactive behaviour in HCI.%

\keywords{Interactive Behaviour Modelling \and Natural Language Processing \and Mouse and Keyboard Input \and Out-of-the-lab Dataset.}
\end{abstract}

\input{introduction}
\input{related_work}
\input{dataset}
\input{method}
\input{experiments}
\input{discussion}
\input{conclusion}

\bibliographystyle{splncs04}
\bibliography{main}

\appendix
\input{IDT}
\input{BPE}
\input{DataAmount}

\end{document}

%% file: introduction.tex
\section{Introduction}
\begin{figure}[t]
  \centering  
  \includegraphics[width=\textwidth]{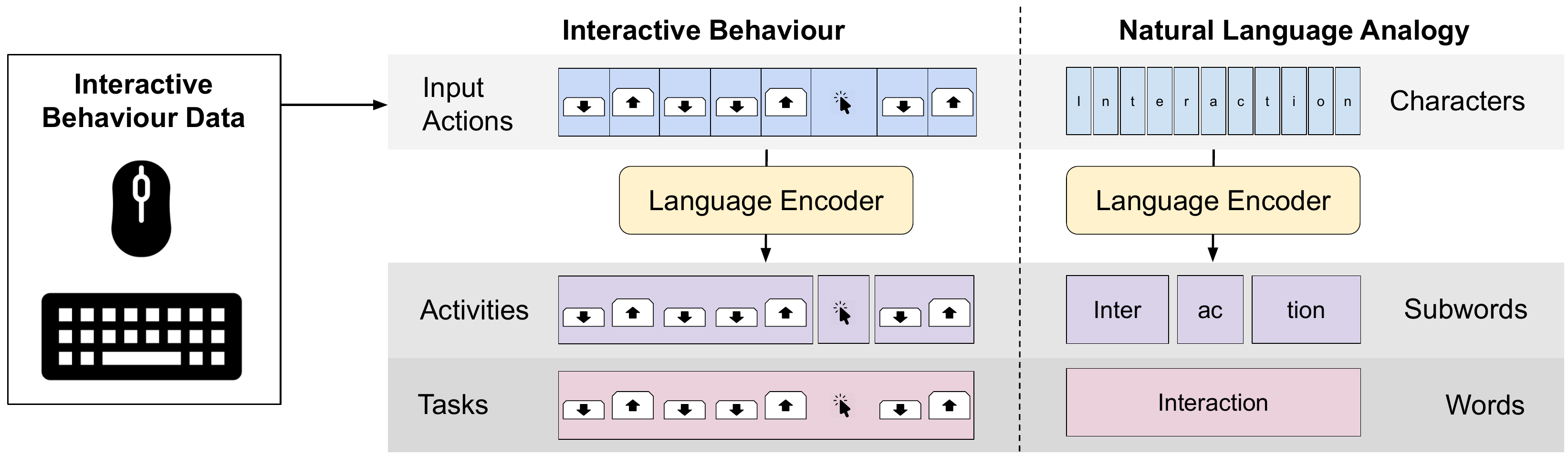}
  \caption{
   Given that both interactive behaviour and natural language are sequential and hierarchical, we explored their similarity by applying an NLP method (a language encoder) to model mouse and keyboard behaviour. %
  }  
  \label{fig:teaser}
\end{figure}
Computational modelling of interactive behaviour has emerged as a key component of intelligent user interfaces (IUIs) in human-computer interaction (HCI)\\\cite{xu2016spatio,zhang2022predicting,antal2021sapimouse,acien2020typenet,dhakal2018observations}.
For example, understanding interactive behaviour helps HCI researchers and user experience (UX) designers analyse and improve interactive systems~\cite{salmeron2014evaluation,bi2012multilingual}.
Mouse and keyboard input is particularly promising %
because it is readily available on a large number of devices and pervasively used in daily life~\cite{xu2016spatio,sun2016shared}.
Interactive behaviour consists of low-level, atomic input actions that cannot be further decomposed~\cite{motwani2015multimodal}, which may resemble characters in natural language.
Furthermore, a sequence of such actions (an activity) that can reflect higher-level interaction goals may resemble a (sub)word that is a sequence of characters with semantic meanings.
As such, interactive behaviour has both a sequential (actions happen one after another) and a hierarchical structure (a sequence of actions forms an activity driven by specific interaction goals), and hence may be similar to natural language (see Fig.~\ref{fig:teaser}).
On the other hand, NLP methods, leveraging the sequential and hierarchical structure of input data, have recently achieved groundbreaking success in various downstream tasks like machine translation and question-answering~\cite{Pennington2014GloVeGV,kim2016character,jawahar2019does,kunchukuttan2016learning}.
However, analysing the possible similarity and link between interactive behaviour and natural language remains under-explored in HCI. %
One notable exception is the work by Han et al.\ that %
encoded $n$ consecutive actions (like mouse clicks) into tokens to learn action embeddings~\cite{han2020modelling}.
However, at its core, the method uses n-gram, which limits the length of action sequences to a fixed length $n$ and requires a dedicated search for its optimal value.
Moreover, the vocabulary size grows exponentially as $n$ increases~\cite{subba2017host}.
Due to such drawback, n-gram has been dropped in NLP in favour of more flexible methods such as byte pair encoding (BPE)~\cite{raffel2020exploring,radford2019language}.
BPE and its variants are used in a significant number of large language models (LLMs) to encode text as subwords, allowing rare or unseen words to be handled without introducing new tokens every time~\cite{sennrich2015neural,schuster2012japanese}.
Additionally, subwords in the vocabulary generated by BPE can have various lengths, allowing a rich and flexible vocabulary.
In this work, we explore the similarity between mouse and keyboard behaviour and natural language, by using BPE to learn a vocabulary, i.e., a set of activities, which is further used to encode the behaviour to perform interactive task recognition.
Knowing which task the user is conducting is essential for adaptive interactive systems that aim to understand interactive behaviour and interaction goals~\cite{pasqual2014mouse,fu2017your,hu2022EHTask}.

Existing mouse and keyboard datasets were typically collected in controlled laboratory settings, although behaviour tends to be more natural in out-of-the-lab settings~\cite{mazilu2015wearable}.
We evaluate the method on two datasets that cover both settings and offer both modalities.
For the lab setting, we chose the Buffalo dataset collected by Sun et al.~\cite{sun2016shared} as it is the largest available dataset~\cite{murphy2017shared}.
For the out-of-the-lab setting, given a lack of suitable publicly available data, we collected a novel multimodal dataset named EMAKI (\underline{E}veryday \underline{M}ouse \underline{A}nd \underline{K}eyboard \underline{I}nteractions)\footnote{\textcolor{\reviewcolor}{The dataset and code are available here: \url{https://git.hcics.simtech.uni-stuttgart.de/public-projects/EMAKI}}}
EMAKI was collected from 39 participants performing three interactive tasks: \textit{text entry and editing}, \textit{image editing} and \textit{questionnaire completion}.
These tasks can be found in a wide range of applications and UIs, and cover varying types of mouse and keyboard actions.

On the two datasets, vocabulary analysis shows that BPE could learn explainable activities, 
e.g., reflecting graphical user interface (GUI) layouts %
and indicating interaction goals such as performing mouse dragging or keyboard shortcuts.
Results from interactive task recognition show that BPE %
outperformed other methods on both modalities and datasets.
In summary, our contributions are three-fold:
    (1) We collect EMAKI, a novel 39-participant out-of-the-lab mouse and keyboard dataset.
    (2) We explore the potential similarity between natural language and mouse and keyboard behaviour by learning meaningful activities via a commonly used NLP method, BPE.
    (3) We show that encoding with BPE also improves the performance of interactive task recognition.
As such, our work uncovers the similarity between natural language and interactive behaviour, showing the potential for applying the new pack of methodology, i.e., NLP methods, to computational interactive behaviour modelling in HCI.

%% file: related_work.tex
\section{Related Work}
\label{sec:relatedwork}

\subsection{Modelling Interactive Behaviour in HCI}
Classical HCI approaches include descriptive models, e.g., Fitts's Law~\cite{accot1997beyond}, and predictive models, e.g., the keystroke-level model (KLM)~\cite{card1980keystroke}. %
However, they are limited in strict controls and modelling simple tasks like pointing to a target or routine tasks that have to be specified step by step~\cite{card1980keystroke}.
Recent research used 1D convolutional neural networks (CNN)~\cite{hu2020dgaze,hu2021FixationNet}, long short-term memory (LSTM)~\cite{hu2022EHTask} and gated recurrent unit (GRU)~\cite{hu2022EHTask} to encode gaze and head behaviour, based on the sequential structure, while others focused on spatial analysis and modelling~\cite{hu2019sgaze,hu2020dgaze}.
Specifically, Xu et al.\ modelled mouse and keyboard behaviour by accumulating cursor positions into binary attention maps~\cite{xu2016spatio}.
Other researchers modelled interactive behaviour from a statistical perspective.
For example, Borji et al.\ used Hidden Markov Models (HMM) to encode motor actions including mouse clicks, mouse positions, and joystick positions in video games~\cite{borji2012probabilistic}, while Sun et al.\ applied Gaussian mixture models (GMM) on keystrokes in text editing tasks~\cite{sun2016shared}.
Researchers also encoded eye movements~\cite{bulling08_pervasive} or gestures~\cite{wang2009human,shirahama2017generality} into strings for activity recognition.
Given that interactive behaviour has a sequential and hierarchical structure, which may resemble natural language, we explored modelling interactive behaviour from an NLP perspective.

\subsection{Encoding Methods for Natural Language}
Recent attractive success in NLP has been largely attributed to methods that efficiently encode characters~\cite{kim2016character}, words~\cite{Pennington2014GloVeGV} or sentences~\cite{Reimers2019SentenceBERTSE} into a vector representation. 
HCI researchers also followed this trend to model GUIs~\cite{Li2021Screen2VecSE,%
Wang2021Screen2WordsAM} or behavioural differences over time~\cite{han2020modelling}.
A key requirement for such methods is %
to encode or tokenise the input to generate a usable vocabulary of concepts. 
Due to the clear structure of natural language, %
NLP methods encode at the character, subword or word level.
One popular approach is n-gram, which uses $n$ words in a sequence to determine the context where commonly \textcolor{\reviewcolor}{$n\leq5$~\cite{han2020modelling,jansen2021next,inoue2016classification,reani2018investigation}}.
However, such a method is limited by the choice of $n$, and the exponential increase of vocabulary size along $n$.
More promising approaches learn a vocabulary of subwords, among which BPE has been widely used given that it allows rich and flexible vocabulary and understanding rare or unseen words~\cite{Wang2020NeuralMT,kudo2018subword,radford2019language,raffel2020exploring}.
Consequently, we employ BPE as the NLP method to create a vocabulary for interactive behaviour.

\subsection{Analysis and Modelling of Mouse and Keyboard Behaviour}
The mouse and keyboard are among the most widely used input modalities in daily interactions with computers~\cite{xu2016spatio,sun2016shared}.
Some researchers only focused on one modality, i.e., mouse or keyboard.
Arapakis et al.\ explored different representations of mouse movements in web search tasks, including time series, heatmaps, and trajectory-based images~\cite{arapakis2020mouse}, while Antal et al.\ employed 1D CNN to encode mouse actions including click and drag~\cite{antal2021sapimouse}.
Dhakal et al.\ analysed keystroke patterns in a transcription typing task by correlation analysis~\cite{dhakal2018observations}, while Acien et al.\ employed LSTM to encode keystroke sequences in free text typing~\cite{acien2020typenet}.
In contrast, Sun et al.\ explored both mouse and keyboard actions in two typing tasks, yet the work was limited to fully controlled laboratory settings~\cite{sun2016shared}. %

%% file: dataset.tex
\section{Datasets for Evaluation}
Although interactive behaviour, and specifically mouse and keyboard data, has been widely studied in HCI~\cite{sun2016shared,xu2016spatio}, most existing datasets have been collected in strictly controlled laboratory settings.
Laboratory settings have the advantages of control and internal validity, but their ecological validity is highly limited~\cite{apaolaza2013understanding}.
Our out-of-the-lab data collection did not control where, when, how long and via which laptop or desktop participants could join, allowing more natural behaviour~\cite{mazilu2015wearable,petersen2021pedagogical}.
In addition, most datasets only include either mouse or keyboard data, while
we opted for evaluations on both modalities. %
As such, we analysed mouse and keyboard behaviour from the in-the-lab Buffalo dataset~\cite{sun2016shared} and EMAKI, a novel multimodal out-of-the-lab dataset that we collected specifically for this purpose, given lacking suitable publicly available data.
To evaluate constraints in data collection from a time perspective, \textit{task} and \textit{study} completion times were calculated.
The former only counts the time spent on tasks, while
the latter refers to finishing the entire study, including pauses. %

\subsection{The Buffalo Dataset}
\label{sec:Buffalo}
To the best of our knowledge, Buffalo~\cite{sun2016shared} is the largest publicly available in-the-lab dataset containing both mouse and keyboard interactions.
The dataset was collected with standalone keyboards over three sessions. %
148 participants performed two typing tasks: transcribing a pre-defined text and typical office activities, such as answering predefined questions and sending emails.
The average number of mouse actions and keystrokes per participant exceeded 19\,K and 17\,K, respectively.
75 participants completed both tasks with the same keyboard, while the remaining used three keyboards across sessions.
Data from the former 75 participants were used in this work for a more controlled condition, following~\cite{xiaofeng2019continuous}.
The average \textit{task} completion time was 41.71\,mins ($\mathrm{SD}=6.34$), while the average \textit{study} completion time was slightly longer, 41.81\,mins ($\mathrm{SD}=6.27$), indicating that participants barely took breaks in this constrained setting.

\subsection{The EMAKI Dataset}
\label{sec:Ourdataset}
We opted for an online study %
including three tasks: text entry and editing, image editing, and questionnaire completion.
These tasks can be found in a wide range of interactive applications and UIs, and cover varying types of mouse and keyboard actions~\cite{xu2016spatio,sun2016shared}.
Furthermore, the tasks are neither limited to a particular real-world application \cite{chuda2014usage,brown2014finding} nor too controlled or artificial %
\cite{dhakal2018observations,zhang2022predicting,zhao2020reading}, different from the typing-focused tasks in Buffalo. %
Two short assessments were designed to analyse if participants show different proficiencies in using mouse and keyboard.

The study was implemented as a web application and hosted on our university server. The link to the study was sent directly to the participants.
The frontend was implemented in JavaScript, while the backend consisted of a Node.js server and an SQLite database. %
We recorded clicks and key presses with separate events for press and release, mouse movements and their associated timestamps.

\paragraph{\textbf{Participants}}
\label{sec:participants}
We recruited 52 participants through university mailing lists and social networks.
12 participants who did not finish the study and one teenage participant were filtered out, leading to \numberParticipants participants in the end (18 female, 18 male and 3 ``other gender'').
Their ages ranged between 18 and 54 years ($\mathrm{M}=25.05, \mathrm{SD}=6.51$).
Participants completed the study from 16 countries.
On average, they reported having used mouse and keyboard for 13.64 years ($\mathrm{SD}=6.80$).
15 participants used laptop touchpads, while the others used traditional mice.
28 participants used laptop keyboards and the rest used standalone keyboards.

\paragraph{\textbf{Interactive Tasks}}

\begin{figure}[t]
    \centering
    \includegraphics[width=\textwidth]{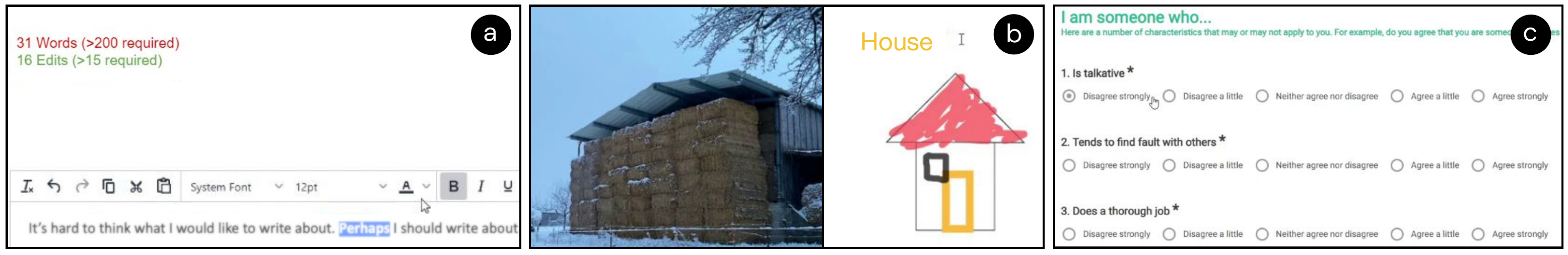}
    \vspace{-0.5cm}
    \caption{Screenshots of the three interactive tasks %
    in our online study: (a) text entry and editing, %
    (b) image editing, %
    and (c) questionnaire completion.%
    }
    \label{fig:main-tasks}
\end{figure}

In \textcolor{\reviewcolor}{task} \textit{text entry and editing}, participants wrote a piece of text in English in a text editor\footnote{\url{https://github.com/tinymce/tinymce}} for one trial (Fig.~\ref{fig:main-tasks}a).
We did not specify the topic but offered suggestions, such as ``summarise a movie/TV series/documentary that you recently watched'' or ``describe your pet''.
We asked participants to write %
$\geq$200 words and apply %
$\geq$15 formatting rules, e.g. change font size or alignment.
\textcolor{\reviewcolor}{We allowed any operation provided by the editor, such as copy-paste and undo.}
Two counters in the top left showed the number of words they already typed and formatting operations they applied.
These counters were initially red and turned green once the minimum thresholds were reached.

In \textcolor{\reviewcolor}{task} \textit{image editing}, participants were presented with two images shown side-by-side in an image editor\footnote{\url{https://github.com/nhn/tui.image-editor}} (Fig.~\ref{fig:main-tasks}b).
The image on the left was a real photograph, whereas the image on the right was a sketch.
\textcolor{\reviewcolor}{On either or both sides, participants performed operations provided by the editor in any order they wanted.
Candidate operations are drawing, cropping, flipping, rotating, adding icons and adding filters.}
To proceed to the next task, they had to perform at least 100 editing operations.
In addition, we asked them to add at least one text box that contained a minimum of 10 characters.
Similarly to the previous task, counters showed the task progress.

\textit{Questionnaire completion} involved participants in completing four questionnaires\footnote{\url{https://github.com/surveyjs/survey-library}}, leading to four trials (Fig.~\ref{fig:main-tasks}c).
These questionnaires served a dual purpose: providing information about participants, which can serve as metadata for future work on the dataset, while at the same time allowing us to record naturalistic mouse and keyboard data.
The first questionnaire focused on demographics and included questions on gender, age, country of origin, country of residence, experience in using mouse and keyboard, and whether participants had any visual impairments. 
Afterwards were %
three widely-used personality questionnaires: BFI-44 (Big Five)\footnote{\url{https://www.ocf.berkeley.edu/~johnlab/bfi.php}}, BIS-11 (Barratt Impulsiveness Scale)\footnote{\url{http://www.impulsivity.org/measurement/bis11}} and BIS-BAS (the Behavioural Inhibition and Approach System)\footnote{\url{https://local.psy.miami.edu/people/faculty/ccarver/availbale-self-report-instruments/bisbas-scales/}}.%

\paragraph{\textbf{Procedure}}
\label{sec:userstudy-procedure}
Before starting with the tasks, participants were asked to carefully read the study goals and task descriptions.
They were then asked whether they were using a mouse or touchpad, and a laptop or standalone keyboard.
To start the study, participants had to click two checkboxes to confirm that (1) they had read and understood the goals of the study, and (2) their data may be published and analysed for research purposes.
Afterwards, participants performed %
tasks in fullscreen.
If %
they left the fullscreen mode during a task, the task was restarted.
We opted for the design to discourage participants from multitasking. %
To reduce potential effects of task order, half of the initial 52 participants performed the text entry and editing task first, followed by the image editing task, while the other half performed %
in the inverse order.
After data filtering, 24 participants did the text task and then image task, while the other 15 in the inverse order. %
We always showed questionnaires at the end, following studies that also collected personality questionnaires~\cite{hoppe2018eye,muller2018detecting}. 
Detailed guidelines for tasks were available to participants throughout the study.
Participants could contact us whenever they had questions, felt uncomfortable or unsure of any task or wanted to withdraw.
Upon completion of the study, participants were shown their results of personality questionnaires as compensation.
No monetary compensation was made.

\paragraph{\textbf{Dataset Statistics}}
\label{sec:taskduration}
The average task completion time was 37.40\,mins ($\mathrm{SD}=13.91$), in which 16.60\,mins ($\mathrm{SD}=8.51$) were spent on text entry and editing, 6.15\,mins ($\mathrm{SD}=3.60$) on image editing, and 9.84\,mins ($\mathrm{SD}=4.48$) on questionnaires.
The average study completion time was significantly longer, 55.33\,mins ($\mathrm{SD}=29.32$).
In total, we collected 1.14\,M mouse actions and 205\,K keyboard actions.
38\% of mouse actions were generated from the image editing task, 43\% from questionnaire completion, while only 19\% came from the text entry and editing task.
Text entry and editing contributed 92\% of the keyboard actions, while only 8\% were from the other two tasks (image editing: 3\%, questionnaire completion: 5\%).

\paragraph{\textbf{Assessments of Proficiency}}
\label{sec:proficiency}

\begin{figure}[t]
    \centering
    \includegraphics[width=\textwidth]{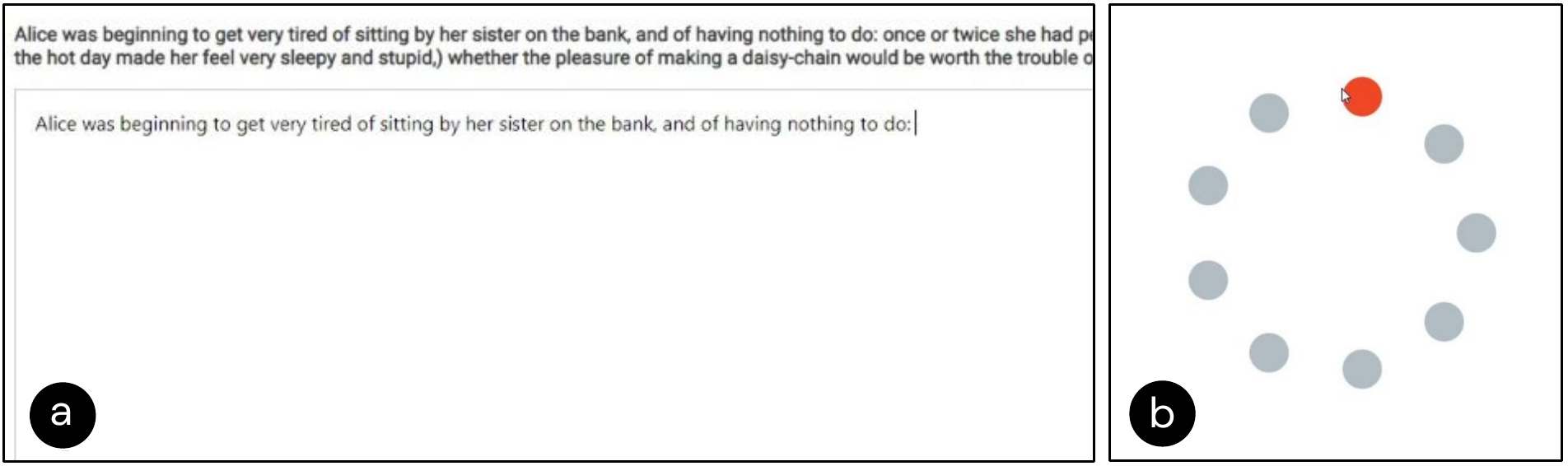}
    \caption{Two proficiency assessments: (a) text typing and (b) move and click.
    }
    \label{fig:calibration-tasks}
\end{figure}

Before interactive tasks, our study also included two short assessments to analyse if participants who used different types of input devices showed different proficiencies in using mouse and keyboard.
The two assessments were \textit{text typing} for keyboard proficiency and \textit{move and click} for mouse proficiency, shown in Fig.~\ref{fig:calibration-tasks}.
\textit{Text typing} involved copying a short piece of text ($\sim$100 words, Fig.~\ref{fig:calibration-tasks}a) as quickly as possible~\cite{grabowski2008internal}.
The average duration of key presses and the number of keys pressed per minute were calculated as keyboard metrics~\cite{grabowski2008internal}.
\textit{Move and click} was inspired by a Fitts's Law task~\cite{soukoreff2004towards}, where participants clicked an orange dot that randomly appeared at a predefined location as quickly as possible over multiple rounds.
Once clicked, the orange dot turned grey and another random dot turned orange (Fig.~\ref{fig:calibration-tasks}b).
Fitts's law~\cite{fitts1954information} models movement time as $MT=a+b \log_{2} \left(\frac{2d}{w}\right)$,
where $d$ is the distance between the centre of the target and the starting point;
$w$ is the width of the target; %
$a$ and $b$ are constants that can be interpreted as the delay and the acceleration.
Based on $d$, $w$ and $MT$ recorded in \textit{move and click}, we computed $a$ and $b$ via linear regression and used them as metrics of mouse proficiency.

Based on the type of mouse (touchpad vs. traditional mouse), we split participants into two groups and then calculated mouse metrics from data collected in the mouse assessment.
A Mann-Whitney U test showed that both metrics were significantly different between the two groups.
One reason is that touchpad and traditional mouse lead to different pointing speeds and accuracies~\cite{hertzum2013effect}.
Then, we split participants into two groups based on using a laptop or standalone keyboard.
No significant difference was found in keyboard metrics calculated from the keyboard assessment.

%% file: method.tex
\section{Modelling Interactive Behaviour with an NLP Method}%
\label{sec:method}

\begin{figure}[t]
    \centering
    \includegraphics[width=\textwidth]{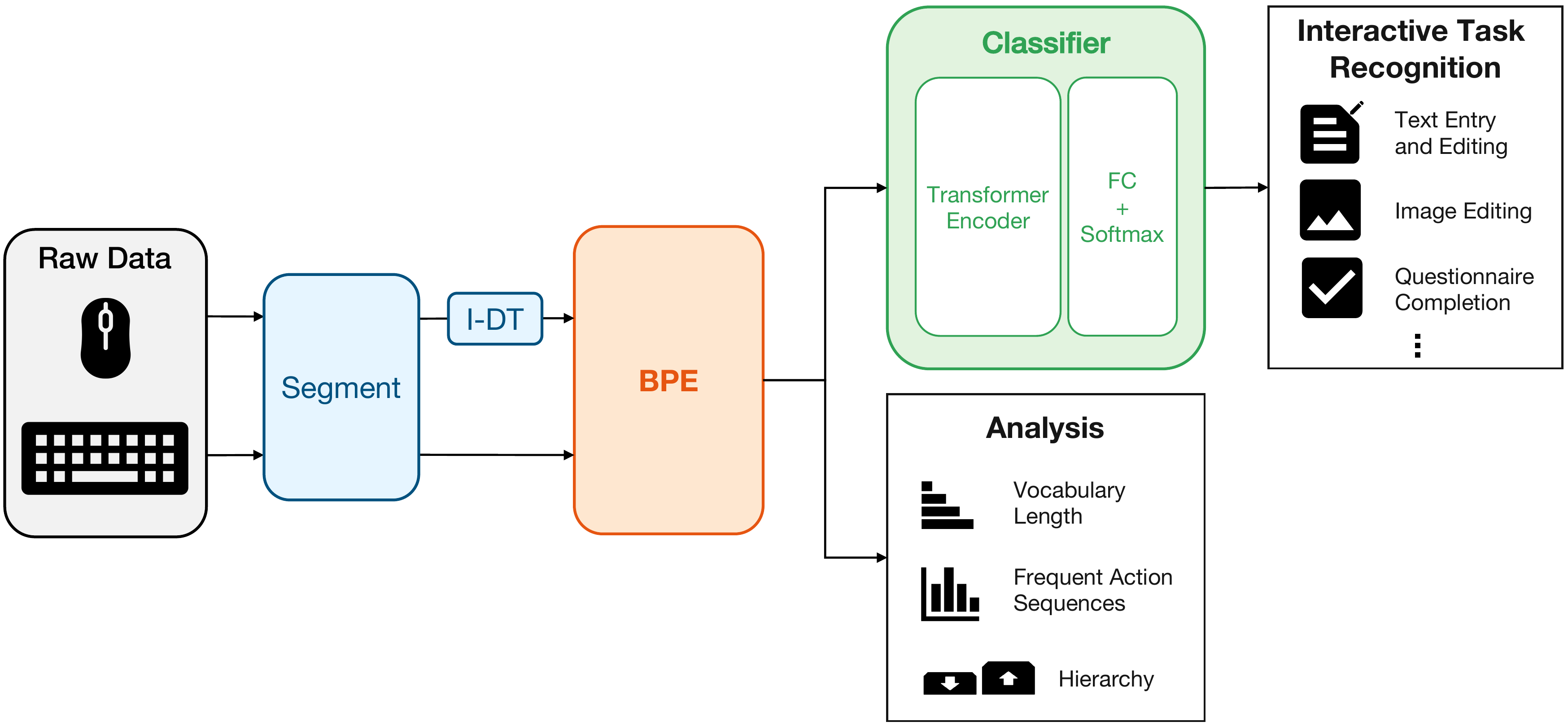}
    \caption{Overview of our pipeline of exploring modelling interactive behaviour from an NLP perspective.
    }
    \label{fig:pipeline}
\end{figure}
As Fig.~\ref{fig:pipeline} shows, the raw data (mouse and keyboard action sequences) are first segmented into subsequences.
Core to our approach is BPE %
learning a vocabulary of subwords, i.e. a set of meaningful mouse and keyboard activities, and then encoding the behaviour based on the vocabulary.
As BPE requires discrete inputs, mouse data are preprocessed additionally using
the dispersion-threshold identification (I-DT) algorithm, that converts continuous-valued mouse coordinates into discrete tokens.
The encodings generated by BPE are then evaluated in two ways to explore if a natural language-like structure exists in mouse and keyboard behaviour that can be captured by this widely used NLP method:
(1) analyse the semantic meaning of the vocabulary, i.e., interaction goals underlying learnt activities, and (2) as input to train a Transformer-based classifier for task recognition.
The two evaluations are demonstrated in Section~\ref{sec:evaluation}.

\subsection{Data Preprocessing}
\label{sec:mousepreprocessing}
Different from natural language where words and sentences are separated by spaces and punctuations, modelling interactive behaviour first requires %
splitting data into smaller units. 
Thus, a sliding non-overlapping window was used to segment the long raw data.
On the keyboard actions, the window lengths $L_{win}$ were empirically set to 10, 50, and 100.
The window lengths $L_{win}$ for the mouse actions were set to 20, 100 and 200, as we observed on both datasets that, the number of generated mouse actions for a fixed time window is roughly twice as many as the keyboard actions.
When using both modalities jointly, the window lengths were set to the mean value of those for single modalities, i.e. $L_{win}=$ 15, 75 and 150.
For keyboard actions, the action type and the key value were concatenated as a token, e.g., \textit{KeyDown\_a} (a${\downarrow}$) or \textit{KeyUp\_Shift} (Shift${\uparrow}$).
Buffalo recorded 91 key values, while EMAKI had 137 values, yielding 182 and 274 atomic actions forming the starting vocabulary, respectively.
With more types of keys, EMAKI can potentially reflect more behaviour varieties.

Participants completed our study on their own computers with different screen resolutions, so we first re-scaled the mouse coordinates to $[0,1]$.
For consistency, we re-scaled Buffalo mouse data to the same range.
We observed two categories of mouse behaviour: \textit{pinpoint}, i.e. interacting with the target UI element in a small area, where moves are shorter, slower and more concentrated, resembling gaze fixations;
and \textit{re-direction} between targets, resembling fast saccadic eye movements between fixations~\cite{salvucci2000identifying}.
Inspired by gaze fixation detection, we used I-DT~\cite{salvucci2000identifying} to preprocess mouse data (see Appendix).
Then we divided the screen equally into four areas  (0: top-left, 1: top-right, 2: bottom-left, 3: bottom-right).
The action type (move or click), mouse behaviour category (pinpoint or re-direction), and the screen area were concatenated as a token, e.g., \textit{Move\_Redirection\_Area0} or \textit{Click\_Pinpoint\_Area3}.
When representing clicks, Buffalo only recorded a \textit{Click}, while we recorded both \textit{Down} (press) and \textit{Up} (release) events.
Therefore, Buffalo has $2{\times}2{\times}4{=}16$ atomic actions and EMAKI has $3{\times}2{\times}4{=}24$.

\subsection{Encoding Mouse and Keyboard Behaviour with BPE}
\label{sec:BPE-method}

We employed BPE (see Appendix for its algorithm) to learn a vocabulary of subwords, i.e., activities that consist of various numbers of consecutive actions.
Starting from the action sequence set $D$, the vocabulary $V$ is built after $k$ iterations.
In each iteration, the most frequent pair of actions or activities form a new activity, which is added into $V$ and used to update $D$.
We consider each action as a character, given it is an inseparable, atomic unit.
The initial vocabulary is composed of actions and one extra token representing the end of the action sequence from one task trial.
Thus, the initial vocabulary sizes are $|V|_\text{mouse}=17$ and $|V|_\text{key}=183$ in Buffalo, and $|V|_\text{mouse}=25$, $|V|_\text{key}=275$ in EMAKI.
We set $k$ to 300, 600 and 900 empirically.%

%% file: experiments.tex
\section{Evaluations of the NLP Method}
\label{sec:evaluation}
As mentioned at the beginning of Section~\ref{sec:method}, BPE was evaluated in two ways: (1) we analysed its vocabulary to examine if the way of learning semantic subwords from characters could learn meaningful activities from interactive actions; and (2) we tested if encoding interactive behaviour in this NLP fashion benefited a downstream task, interactive task recognition, using a Transformer-based classifier.
\subsection{Analysis of the Learnt Vocabulary}
\label{sec:experiment-vocabulary}
We first examined statistics of the vocabulary including its size and activity lengths.
Then we analysed semantic meanings of the most frequent and long activities.
Frequent activities are short, low-level and pervasively exist in various activities, while long activities reflect high-level and complex goals.

\subsubsection{Vocabulary Statistics.}

\begin{figure}[t]
    \centering
    \vspace{-1cm}
    \includegraphics[width=\linewidth]{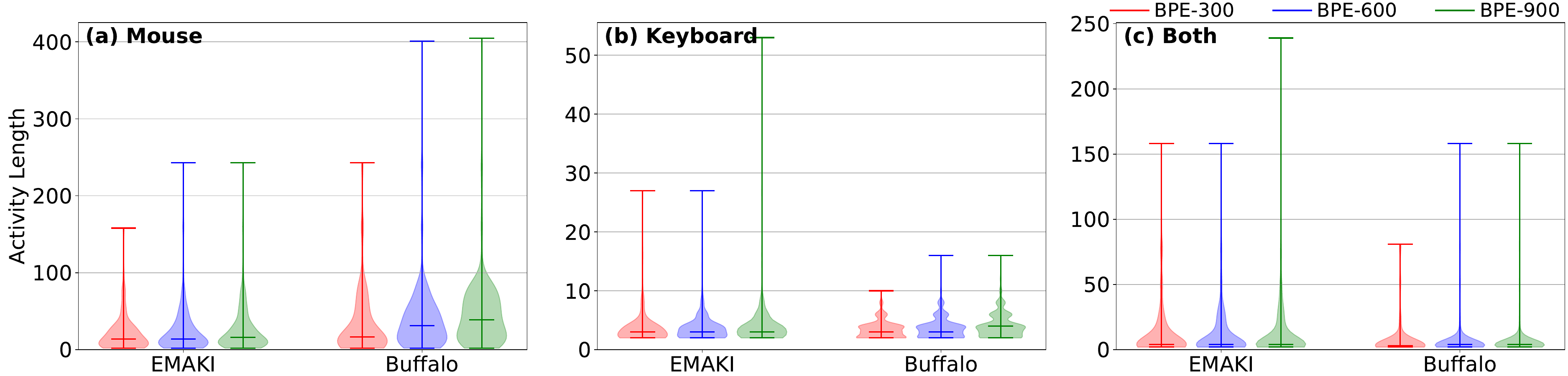}
    \vspace{-0.5cm}
    \caption{Violin plots for the lengths of activities learnt by BPE after 300 (in red), 600 (in blue) and 900 (in green) iterations, of (a) mouse, (b) keyboard and (c) both modalities on EMAKI and Buffalo datasets. Each bar shows the range of activity lengths, while the middle line indicates the median length. The y-axes are scaled according to the range in each subplot.}
    \label{fig:violin}
\end{figure}

As Fig.~\ref{fig:violin}a shows, in EMAKI the maximum length of mouse activities reached 243 actions (BPE-900), while the median length was 16.
The longest keyboard activity had 53 actions, while the median length was 3 (Fig.~\ref{fig:violin}b).
When using both modalities jointly, the maximum activity length was 239 after 900 iterations, while the median length was 4 (Fig.~\ref{fig:violin}c).
In Buffalo, the lengths of activities had a maximum of 405 and a median of 39 from mouse behaviour (Fig.~\ref{fig:violin}a); a maximum of 16 and a median of 4 from keyboard behaviour (Fig.~\ref{fig:violin}b); and a maximum of 158 and a median of 4 from joint modalities (Fig.~\ref{fig:violin}c).
Mouse activities were longer than keyboard activities, indicating that the preprocessed mouse data were more similar compared to preprocessed keyboard data.
Comparisons between datasets show that mouse activities in EMAKI were more diverse, while Buffalo contained more diverse keyboard activities.

\begin{table}[t]
    \vspace{-1mm}
    \centering
    \begin{tabular}{|l|c|c|c|c|c|c|} 
        \hline
        Dataset & \multicolumn{3}{c|}{EMAKI} & \multicolumn{3}{c|}{Buffalo}\\
        \hline
        Method & BPE-300 & BPE-600 & BPE-900 & BPE-300 & BPE-600 & BPE-900 \\
        \hline
         Mouse & 322 & 622 & 921 & 310 & 609 & 909\\
        Keyboard & 513 & 808 & 1103 & 473 & 770 & 1067\\
        Both & 569 & 864 & 1163 & 496 & 790 & 1084\\
        \hline
    \end{tabular}    
    \vspace{2mm}
    \caption{Vocabulary sizes generated using BPE after 300, 600 and 900 iterations, on EMAKI and Buffalo datasets.}
    \label{tab:vocabsize}
    \vspace{-4mm}
\end{table}

Table \ref{tab:vocabsize} shows the vocabulary sizes generated by BPE on the two datasets.
Note that starting from BPE-$k$ and running the algorithm for $k$ more iterations, the vocabulary size increases by approximately $k$ elements -- showing that BPE overcomes the issue of exponential growth of vocabulary size in n-gram.

\subsubsection{Frequent Activities.}
\begin{table}[t]
    \centering
    \begin{tabular}{|c|c|c|c|c|c|c|c|c|c|c|}
        \hline
        Rank & 1 & 2 &3&4&5&6&7&8&9&10 \\
        \hline
        EMAKI    & $\sqcup{\downarrow}$, $\sqcup{\uparrow}$
        & ${\Leftarrow}{\downarrow}{,}{\Leftarrow}{\uparrow}$
        & e${\downarrow}$, e${\uparrow}$
        & t${\downarrow}$, t${\uparrow}$
        & a${\downarrow}$, a${\uparrow}$
        & o${\downarrow}$, o${\uparrow}$
        & ${\Leftarrow}{\downarrow}{,}{\Leftarrow}{\uparrow}{,}{\Leftarrow}{\downarrow}{,}{\Leftarrow}{\uparrow}$
        & i${\downarrow}$, i${\uparrow}$
        & s${\downarrow}$, s${\uparrow}$
        & n${\downarrow}$, n${\uparrow}$\\
        \hline
        Buffalo & $\sqcup{\downarrow}$, $\sqcup{\uparrow}$
        & ${\Leftarrow}{\downarrow}{,}{\Leftarrow}{\uparrow}$
        & e${\downarrow}$, e${\uparrow}$
        & t${\downarrow}$, t${\uparrow}$
        & o${\downarrow}$, o${\uparrow}$
        & i${\downarrow}$, i${\uparrow}$
        & a${\downarrow}$, a${\uparrow}$
        & s${\downarrow}$, s${\uparrow}$
        & n${\downarrow}$, n${\uparrow}$
        & l${\downarrow}$, l${\uparrow}$\\
        \hline
    \end{tabular}
    \caption{The ten most frequent keyboard activities found by BPE. The $\sqcup$ symbol represents Space. The $\Leftarrow$ symbol means Backspace. The down arrow $\downarrow$ and up arrow $\uparrow$ denote \textit{KeyDown} and \textit{KeyUp}, respectively.}
    \label{tab:top10key}
\end{table}

The three BPE iterations learnt the same top-10 frequent keyboard action sequences as shown in Table~\ref{tab:top10key}.
Eight out of ten action sequences are the same on the two datasets, although they were collected from different participants in different experimental settings, indicating that generalised patterns underlie keyboard behaviour.
The interaction goal behind the most frequent activity is to press the spacebar, which is in line with the observation that spaces occur often when typing in various languages.
The second frequent activity reflects an intention of pressing Backspace which is frequently and widely used to %
correct what has been typed.
Most frequent activities correspond to character keystrokes, and reflect the top-7 most frequent English letters: ``e" (12.15\%),``a" (8.67\%), ``t" (8.60\%), ``i" (7.53\%), ``o" (7.38\%), ``n" (7.34\%) and ``s" (6.63\%)~\cite{grigas2018letter}.
The difference in their order may be due to that the datasets are limited to specific typing scenarios and not representative of the entire English language.
We also noticed that the left and right arrows, for redirecting typing locations, were also frequent on both datasets.

The most frequent ten mouse action sequences learnt by BPE were also the same on the two datasets.
All of them are mouse moves of pinpoint, implying that participants follow similar ways to interact with UI targets even in different tasks and settings.
These pinpointing regions were primarily in the top-left and bottom-left areas, while fewer pinpoints fell on the right side.
This matches the layouts of not only general GUIs but also those used in our user study.
For example, menu bars and sidebars are commonly at the top and to the left of interactive windows, respectively.
Also, our text formatting tools were at the top of the text editor.
The image editing tools were in the leftmost of the image editor.
Additionally, our questionnaires were left-aligned, so the choices for participants to click lay to the left.

\subsubsection{Interaction Goals behind Activities.}
We also analysed long activities to examine if BPE learnt a hierarchy, i.e., if atomic actions form meaningful activities driven by complex goals.
An example is ``Dot${\downarrow}$, Dot${\uparrow}$, Space${\downarrow}$, Space${\uparrow}$, Shift${\downarrow}$, i${\downarrow}$, i${\uparrow}$, Shift${\uparrow}$, Space${\downarrow}$, Space${\uparrow}$''.
The goal behind the whole sequence is to start a sentence with the word ``I'', in line with the common texting or typing scenario of writing about oneself.
It consisted of the low-level goal of pressing each aforementioned key, which was further composed of atomic actions \textit{KeyDown} and \textit{KeyUp}.
BPE also learnt ``Space${\downarrow}$, Space${\uparrow}$, Backspace${\downarrow}$, Backspace${\uparrow}$'' from EMAKI, suggesting that participants typed at a faster pace than their thought process.
Another example is ``Ctrl${\downarrow}$, s${\downarrow}$, s${\uparrow}$, Ctrl${\uparrow}$'' from Buffalo, representing the shortcut for saving files.
Looking at mouse behaviour, BPE captured drag behaviour, represented as a \textit{MouseDown} action followed by multiple \textit{MouseMove} actions and ending with a \textit{MouseUp} action.
Another learnt long activity %
had 37 actions %
with 35 moves and a click as pinpoint in area 0, reflecting the goal of adjusting the cursor to a target and then clicking.

\subsection{Interactive Task Recognition}
\label{sec:task-recognition}
We also evaluated the practical effectiveness of our approach on interactive task recognition.
Knowing which task a user is performing enables adaptive UIs to understand the interactive behaviour and goals~\cite{hu2022EHTask,hu21_User}.
We compared our approach with two baselines: an ablated version which bypasses encoding (noted as NoEncoding) and replacing BPE with an autoencoder (AE).
Autoencoder, consisting of an encoder and a decoder, is trained in a self-supervised way to reconstruct the input with the lowest error.
Therefore, it needs no annotations and has a high generalisability, %
also used on language data~\cite{li2015hierarchical}.
To control variables, i.e., restrict the comparison to the encoding, we set two rules:
(1) to reduce the impact of sophisticated designs of the encoders, use vanilla AE and BPE; (2) use the same hyperparameter sets for the classifier.

We implemented an AE that includes four components: an embedding layer of dimension $d_{e}=128$ to handle discrete tokens; an encoder component composed of one to three fully connected (FC) layers with hidden dimensions $(64)$, $(64, 32)$ and $(64, 32, 16)$; a decoder component, which is symmetric to the encoder; and a reconstruction component consisting of an FC layer %
and a softmax layer. %
Dropout was added after FC layers to avoid overfitting.
We denote the autoencoder that has one, two, or three FC layers in the encoder and decoder components as AE-1, AE-2 and AE-3.
Cross entropy between the reconstructed sequences and the input was used as the loss function.
After training, the encoder component was used to encode interactive behaviour.

Our task classifier is based on a Transformer~\cite{vaswani2017attention}, which is well known for its success in NLP and capability to handle long dependencies in temporal signals.
The classifier is composed of $N=\{2, 4, 6\}$ Transformer encoder layers, then an FC and softmax layer.
Each Transformer encoder layer had $h=4$ attention heads, $d_\text{model}=\{16, 64\}$ expected features, $d_\text{ff}=4d_\text{model}$ dimension in feedforward layers and uses the ReLU activation function.
During training, we applied label smoothing with $\epsilon=0.1$~\cite{vaswani2017attention}.
We used AdamW optimizer with learning rate $lr=\{10^{-3}, 10^{-4}\}$ and $\beta=(0.9, 0.999)$~\cite{bruckner2022learning} and the cross entropy as loss function.
The training was done on a Tesla V100 GPU with a batch size of 64 and a dropout rate of 0.5.
The classifier was trained for 30 epochs, while the AE was trained for 10 epochs because of its faster convergence.
Because activities in the flexible vocabulary learnt by BPE have different lengths, we padded short samples %
and applied padding masks.

\begin{figure}[t]
    \centering
    \includegraphics[width=\linewidth]{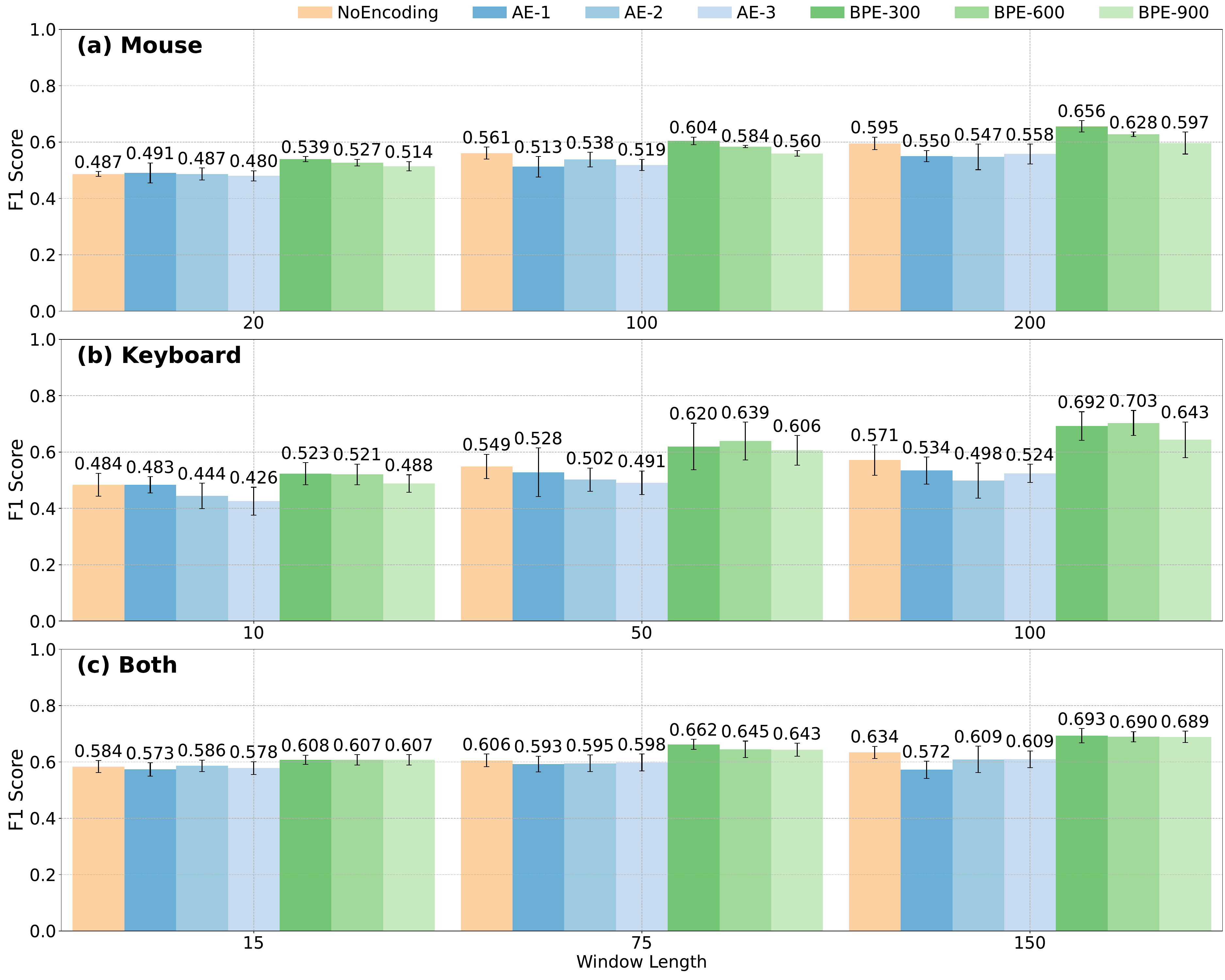}
    \caption{F1 scores of recognising three interactive tasks on EMAKI from (a) mouse, (b) keyboard and (c) both modalities, segmented by different windows.
    Error bars represent the standard deviation from a 5-fold cross-validation.}
    \label{fig:TR-Ours}
\end{figure}

\begin{figure}[t]
    \includegraphics[width=\linewidth]{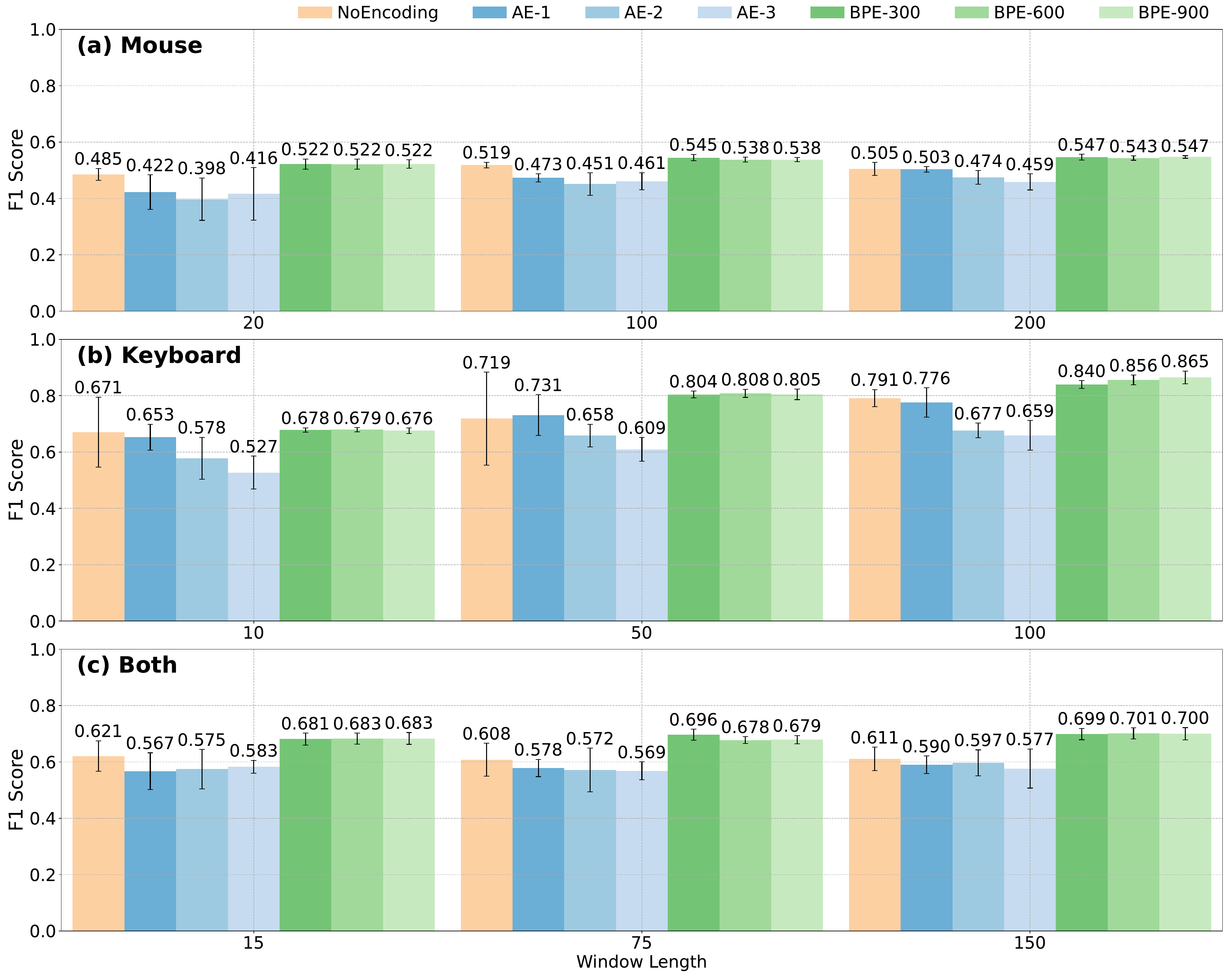}
    \caption{F1 scores of recognising two interactive tasks on Buffalo from (a) mouse, (b) keyboard and (c) both modalities, segmented by different windows.
    Error bars represent the standard deviation from a 5-fold cross-validation.}
    \label{fig:TR-UB}
\end{figure}

EMAKI has three main interactive tasks, posing a three-class classification problem, while Buffalo has two tasks, posing a binary classification problem.
The evaluation follows 5-fold participant-independent cross-validation, where data from 80\% of participants form the training set and the remaining participants form the test set.
This scheme can evaluate the performance of unseen users. %
Macro F1 score~\cite{hoppe2018eye} was chosen as evaluation metric because of the imbalanced classes, %
e.g., most keyboard data were from the text task on EMAKI.
For each model, we report the highest F1 score achieved among all the parameter sets.
Results show that on both datasets methods using BPE encoding outperformed the others (see Fig.~\ref{fig:TR-Ours} and~\ref{fig:TR-UB}).

\paragraph{\textbf{Results on EMAKI}}
\label{sec:TR-Ours}

On mouse data, BPE-300 consistently outperformed other methods (Fig.~\ref{fig:TR-Ours}a).
A one-way ANOVA test showed that differences between methods are significant ($p{<}.001$): $F{=}9.697$ on $L_{win}{=}200$, $F{=}12.396$ on $L_{win}{=}100$ and $F{=}7.194$ on $L_{win}{=}20$. A post-hoc Tukey HSD test further confirmed that BPE-300 significantly outperformed the other methods on $L_{win}{=}200$, $L_{win}{=}100$ ($p{<}.001$ for AE and $p{<}.05$ for NoEncoding) and $L_{win}{=}20$ ($p{<}.01$ for both AE and NoEncoding). 
Fig.~\ref{fig:TR-Ours}b shows that BPE-600 achieved the best results for $L_{win}{=}100$ and $L_{win}{=}50$, whereas when $L_{win}{=}10$ the best was BPE-300.
Differences between methods are significant ($F{=}13.044$, $p{<}.001$ for $L_{win}{=}100$, $F{=}4.620$, $p{<}.01$ for $L_{win}{=}50$ and $F{=}4.220$, $p{<}.01$ for $L_{win}{=}10$). 
Post-hoc Tukey HSD tests confirmed that BPE-600 significantly outperformed NoEncoding ($p{<}.01$) and AE-1 ($p{<}.001$) for $L_{win}{=}100$. %
On joint modalities, BPE-300 performed the best, with %
the highest F1 score of 0.693 (Fig.~\ref{fig:TR-Ours}c). %
Differences between methods were again significant with $F{=}13.996$, $p{<}.001$ on $L_{win}{=}150$, $F{=}5.678$, $p{<}.001$ on $L_{win}{=}75$ and $F{=}2.665$, $p{<}.05$ on $L_{win}{=}15$.
Tukey HSD test indicated that BPE-300 significantly outperformed AE ($p{<}.01$) and NoEncoding ($p{<}.05$) on $L_{win}{=}150$ and both of them at $p{<}.05$ when $L_{win}{=}75$.

In Section~\ref{sec:proficiency}, we report that participants using touchpads and traditional mice show different proficiencies. Therefore, we analysed if such differences affected task
recognition.
We separately performed 5-fold cross-validation based on the two groups.
Since 24 participants used traditional mice while only 15 used touchpads, we randomly selected 15 traditional mouse users to reduce the influence of data amount on performance.
Because BPE-300 on the longest window achieved the best results on mouse data (Fig.~\ref{fig:TR-Ours}a), we used the same setting and did a Mann-Whitney U test on F1 scores achieved from two groups.
To mitigate the randomisation introduced by participant selection, we repeated the above procedure five times.
None of the five tests found a significant difference in performance.
The reason may be that our method does not explicitly encode time information, thus ignoring the speed difference in moving the cursor~\cite{hertzum2013effect}.

\paragraph{\textbf{Results on Buffalo}}
\label{sec:TR-UB}
On mouse data (Fig.~\ref{fig:TR-UB}a), BPE-300 performed the best %
and got the highest F1 score of 0.547. %
One-way ANOVA showed that differences between methods were significant ($p{<}.001$) with $F{=}20.345$ for $L_{win}{=}200$, $F{=}18.609$ for $L_{win}{=}100$ and $F{=}5.589$ for $L_{win}{=}20$).
Post-hoc Tukey HSD tests showed that BPE-300 significantly outperformed NoEncoding ($p{<}.05$) and AE-1 ($p{<}.001$) when $L_{win}{=}200$.
On keyboard data (Fig.~\ref{fig:TR-UB}b), BPE-900 and BPE-600 outperformed other methods. %
Differences between methods are significant with $F{=}30.218$ for $L_{win}{=}100$, $F{=}5.884$ for $L_{win}{=}50$ (both $p{<}.001$) and $F{=}4.791$, $p{<}.01$ for $L_{win}{=}10$.
According to post-hoc Tukey HSD tests, BPE-900 significantly outperformed AE-1 ($p{<}.01$) and NoEncoding ($p{<}.05$) when $L_{win}{=}100$, and BPE-600 significantly outperformed AE-1 ($p{<}.001$) when $L_{win}{=}50$.
On joint modalities (Fig.~\ref{fig:TR-UB}c), BPE resulted in similar %
yet higher F1 scores than baselines.
The best result was achieved by BPE-600 on the longest window %
of 0.701. %
Differences between methods were again significant ($p{<}.001$): %
$F{=}10.733$ for $L_{win}{=}150$; $F{=}11.151$ for $L_{win}{=}75$; and $F{=}7.397$ for $L_{win}{=}15$.
Tukey HSD test showed that BPE-600 significantly outperformed AE-2 ($p{<}.01$) and NoEncoding ($p{<}.05$) on $L_{win}{=}150$; BPE-300 outperformed AE-1 ($p{<}.01$) and NoEncoding ($p{<}.05$) on $L_{win}{=}75$; and BPE-600 outperformed AE-3 ($p{<}.05$) on $L_{win}{=}15$.

\begin{figure}[t]
    \centering
    \includegraphics[width=0.7\linewidth]{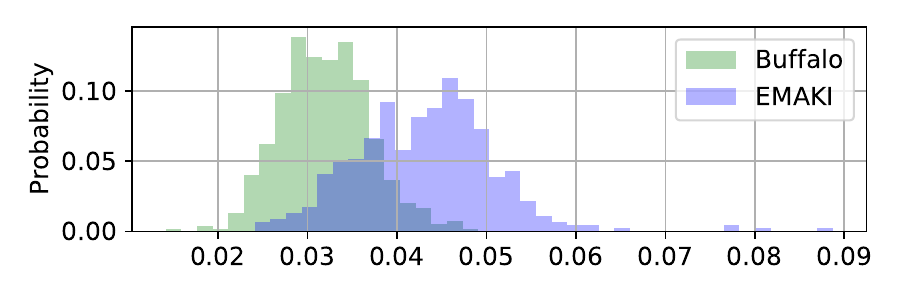}
    \caption{Distribution of the average Euclidean distances between mouse trajectories from different interactive tasks on the two datasets. Smaller distances mean that trajectories from different tasks are more similar.}
    \label{fig:distributions}
\end{figure}

It is noticeable that results obtained from Buffalo mouse data slightly exceeded the chance level and were much worse than those from keyboard data.
A possible reason is that the mouse behaviour on the Buffalo dataset was similar across different tasks.
To verify this, we calculated the average distances between mouse trajectories in different interactive tasks, following~\cite{wulff2019mouse}:
(1) all the mouse actions generated in one trial by one participant were considered one trajectory, on which 101 points were sampled uniformly;
(2) the distance between two trials was defined as the average Euclidean distance between each pair of points on two trajectories;
(3) the distance between two tasks was computed as the average distance between each trial from task 1 and each from task 2.
Fig.~\ref{fig:distributions} shows that the distance between tasks from Buffalo is smaller than EMAKI, suggesting that mouse behaviour generated from the two tasks from Buffalo is similar, consistent with the statistics of BPE vocabulary (Section~\ref{sec:experiment-vocabulary}).
Therefore, it is more difficult to classify tasks based on Buffalo mouse data.

%% file: discussion.tex
\section{Discussion}
\label{sec:discussion}

\subsection{Modelling Interactive Behaviour from a Natural Language Perspective}
\label{sec:discussion-language}
Our work is among the first to explore the similarity between interactive behaviour and natural language, given that both have a sequential and hierarchical structure.
Towards this goal, we applied BPE, which has been commonly used in state-of-the-art large language models to encode mouse and keyboard behaviour.
At the lowest level, input actions were considered as %
characters since they are atomic and inseparable.
For higher levels, BPE learned ``subwords'' from interactive behaviour, which were interactive activities, i.e., action sequences driven by underlying interaction goals.
The analysis of the learnt vocabulary showed that following the same way of learning the semantic hierarchy of language, BPE was able to %
capture meaningful activities %
such as mouse drags, keyboard shortcuts and precisely adjusting the mouse to click on a UI element (Section~\ref{sec:experiment-vocabulary}).
Despite representing just a first exploration, the insights from our analysis underline the similarity between interactive behaviour and natural language, %
and indicate the possibility of applying more powerful NLP methods like BERT~\cite{jawahar2019does,liu2019roberta} to encode interactive behaviour.
Besides the state-of-the-art performances achieved, such LLMs also have noticeable advantages of generalisability and reusability.
They can be pretrained on one dataset and re-used to encode other datasets to solve various downstream tasks with fine-tuning, which is more cost-effective than dedicating a specific large model towards each dataset or task~\cite{sun2019fine}.
Future HCI research can follow such NLP methods to build reusable pretrained interactive behaviour models for better generalisability and cost-effectiveness.

\subsection{NLP Encoding for Interactive Task Recognition}
\label{sec:discussion-compareMK}
Interactive task recognition is one of the key requirements of intelligent interactive systems to understand and adapt to interactive behaviour and interaction goals~\cite{pasqual2014mouse,fu2017your,gajos2004supple,joachims2002optimizing}.
On this recognition task, encoding with BPE significantly outperformed baselines on both datasets, all the modalities and windows. %
Specifically, on our out-of-the-lab, newly collected EMAKI dataset, %
encoding with BPE obtained the highest F1 score of 0.703 recognising three tasks
(Fig.~\ref{fig:TR-Ours}).
On average, BPE improved the F1 score by 0.087 on keyboard data, 0.051 on mouse, and 0.044 on the joint modalities.
On the Buffalo dataset, %
BPE achieved the highest F1 score of 0.865 (Fig.~\ref{fig:TR-UB}) recognising two tasks.
On average, BPE improved the F1 score by 0.080 on joint modalities, 0.053 on keyboard data and 0.035 on mouse data.
These results, from a practical perspective, further reveal the promising effectiveness of modelling interactive behaviour as natural language.

We observed that methods generally achieved better results on longer windows, which may be due to that more actions may uncover richer characteristics of the tasks.
However, increasing the window size yields fewer training samples and makes the recognition model wait longer for a complete window of actions to provide a prediction.
In our experiments, windows that led to the best performance on mouse, keyboard and joint modalities had 200, 100 and 150 actions, respectively.
These values can be a reference for future mouse and keyboard behaviour modelling methods.

In addition, on both datasets, using BPE on keyboard behaviour improved the F1 score more than on mouse behaviour, indicating its better ability of handling keyboard than mouse behaviour.
This finding is expected, as typing on a keyboard is directly linked to expressing natural language. %
A second reason might be that discretising mouse data caused a loss of information\textcolor{\reviewcolor}{~\cite{yue2022ts2vec,zerveas2021transformer}}.
On the joint modalities, we observed a general performance improvement from individual modalities on EMAKI, but not on Buffalo.
As shown in Section~\ref{sec:TR-UB}, Buffalo lacks the diversity in mouse behaviour and thus %
performance achieved by combining mouse and keyboard is in-between that of individual modality.

\subsection{EMAKI Dataset}
\label{sec:EMAKI-discussion}
Most publicly available mouse and keyboard datasets were collected in constrained laboratory settings, such as the Buffalo dataset.
In contrast, our EMAKI is a step towards fully unconstrained settings to allow more natural interactive behaviour.
Our study did not control where, when, or how long participants joined the study.
In addition, participants used their own devices, which contributes to ecological validity.
Consequently, our participant pool is more diverse given that participants are from different countries, and used different input devices and screen resolutions.
All of Buffalo's participants were university students between 20-30 years old, while ours were between 18-54 and covered non-student participants.
Moreover, %
our participants spent various time on tasks as they were freer to pause and resume (as shown in Section~\ref{sec:Buffalo} and \ref{sec:Ourdataset}).
Buffalo primarily uses typing-focused tasks%
, while EMAKI has complementary characteristics and tasks -- like image editing and questionnaires -- encouraging diverse mouse behaviour, as confirmed by our analysis in Section \ref{sec:experiment-vocabulary}.
Furthermore, higher diversity in behaviour can lead to better task recognition performance (Section \ref{sec:task-recognition}).
We also verified that the amount of data in EMAKI is sufficient for training the method for task recognition (see Appendix).
Besides serving as a benchmark for task recognition, the questionnaires included in EMAKI also encourage future research on the interplay between multimodal behaviour and personality traits~\cite{zhao2020reading}.

\subsection{Limitations and Future Work}
Our user study %
covered %
diverse but predefined tasks and did not allow multitasking. %
In the future, we will move towards fully uncontrolled settings. %
Time information may further improve behaviour modelling~\cite{azcarraga2011use,freihaut2021does} and will be explicitly encoded in future work. %
\textcolor{\reviewcolor}{We chose BPE over N-gram due to its flexibility, yet for systems where activities have similar lengths, N-gram might be efficient enough.
An interesting future work is to explore the boundary of where the methods lead over the other.}
Also, even used on both modalities jointly, BPE learned activities composed of single modalities.
A possible reason is that the behaviour of switching between mouse and keyboard is diverse, which BPE could not capture.
Future work can explore %
the use of other %
NLP methods to better learn the interplay between mouse and keyboard behaviour~\cite{liu2019roberta,raffel2020exploring,radford2019language}.
\textcolor{\reviewcolor}{Automatic interpreters can be studied to identify meaningful and interesting insights into behaviour from the BPE vocabulary, instead of human interpretation.}
Moreover, we intend to study other interactive modalities, such as screen touch and mid-air gesture, as well as other HCI downstream tasks like personality recognition.

%% file: conclusion.tex
\section{Conclusion}
We explored the similarity between interactive behaviour and natural language, given that both of them have a sequential and hierarchical structure.
Towards the goal, we applied a widely used NLP method, BPE, to encode mouse and keyboard behaviour by learning its subwords, i.e., activities.
Results on an existing controlled dataset and a novel out-of-the-lab dataset showed that the method can capture meaningful activities.
Moreover, encoding with BPE %
significantly improved interactive task recognition, which is commonly required in intelligent interactive systems. %
Taken together, our exploratory work links interactive behaviour with natural language and provides a promising NLP perspective for modelling interactive behaviour, %
which has the potential to improve the generalisability of computational interactive behaviour models (Section~\ref{sec:discussion-language}) and also performances of interactive behaviour-based HCI tasks.

%% file: IDT.tex
\section{Preprocessing Mouse Data with I-DT}
As written in Section \ref{sec:mousepreprocessing}, the Dispersion-threshold identification (I-DT) algorithm~\cite{salvucci2000identifying} was used to categorise mouse behaviour to \textit{pinpointing} a target (resembling gaze fixations) and \textit{re-direction} between targets.
I-DT operates on a window of duration-threshold consecutive samples.
On this window, it calculates the dispersion value as 
$Dispersion=[max(x)-min(x)]+[max(y)-min(y)]$.
If the dispersion value exceeds the dispersion threshold, samples inside the window are not considered to belong to a pinpoint and the window is slid forward by one sample.
If the value is below the threshold, the samples within the window are considered to belong to a pinpoint.
The window then expands to incorporate new samples until the dispersion value is above the threshold again.
We empirically set the duration threshold to 100\,ms and the dispersion threshold to 0.1.

%% file: BPE.tex
\section{The Algorithm of Byte Pair Encoding}
\begin{algorithm}[t]
    \caption{Byte pair encoding (BPE)~\cite{zhan2019effective,heinzerling2017bpemb}
    }
    \label{alg:BPE}
    \begin{algorithmic}
        \State \textbf{Input:} action sequence set $D$, the number of iterations $k$
        \Procedure{BPE}{$D$, $k$}
        \State $V \gets \text{all unique actions in } {D} $
        \For{$i \gets 1$ to $k$}  %
        \State $t_L, t_R \gets \text{Most frequent two consecutive units (actions or activities) in } {D} $
        \State $t_\text{new} \gets t_L + t_R$\Comment{Merge to form a new activity}
        \State $V \gets V + [t_\text{new}]$
        \State $\text{Replace each occurrence of } t_L, t_R \text{ with } t_{\text{new}} \text{ in } {D}$
        \EndFor
        \State \textbf{return} $V$ 
        \EndProcedure
    \end{algorithmic}
\end{algorithm}
Algorithm \ref{alg:BPE} shows how byte pair encoding (BPE) constructs the vocabulary $V$, as introduced in Section \ref{sec:BPE-method}.

%% file: DataAmount.tex
\section{Analysis of EMAKI Data Amount for Interactive Task Recognition}
\begin{figure}[t]
    \centering
    \includegraphics[width=.5\textwidth]{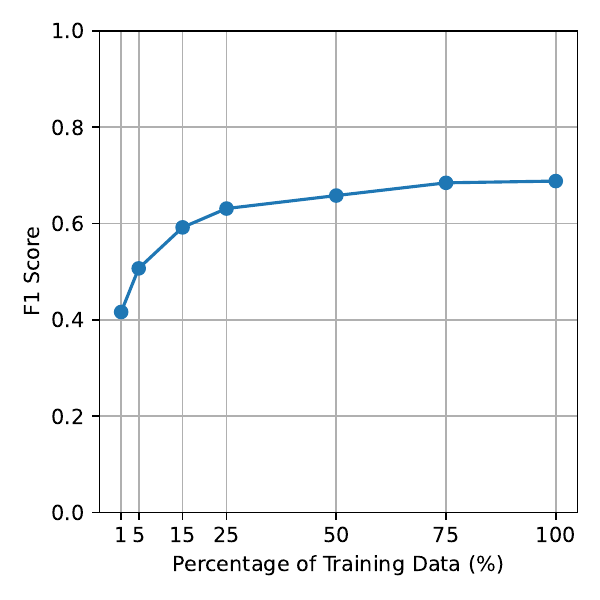}
    \caption{F1 scores of interactive task recognition achieved by BPE, trained on different percentages of the training set.}
    \label{fig:dataamount}
\end{figure}

As written in Section \ref{sec:EMAKI-discussion}, we evaluated if the size of EMAKI allows our data-driven method to recognise interactive tasks.
We used different percentages of the training set to train the method and examined their performances.
According to Figure~\ref{fig:TR-Ours}c, the best results were achieved by BPE-300 when windows have 150 actions.
Therefore, we followed the above setting.
Fig.~\ref{fig:dataamount} shows the results of interactive task recognition by training the model with 1\%, 5\%, 15\%, 25\%, 50\%, 75\% of randomly selected training instances, as well as with the entire training set (100\%).
It can be seen that as the percentage increases, the F1 score first increases fast (before 25\%) but then slowly (25\% to 75\%).
The increase in F1 score from using 75\% of training data and the entire training set was subtle (only 0.004).
Taken together, the amount of data in our dataset is sufficient to perform interactive task recognition.
